\documentclass[aip,amsmath,amssymb,preprint]{revtex4-1}
 \usepackage{graphicx}
 \usepackage{dcolumn}
\usepackage{bm}
\usepackage[T1]{fontenc}
\usepackage{mathptmx}

 \usepackage{natbib}
\usepackage{subcaption}
 \usepackage{amsmath}
 \usepackage{textcomp}
 \usepackage{float}
 \usepackage{rotating}
 \usepackage{stackrel}
\begin{document}
 
\title{Resonance enhanced two-photon cavity ring-down spectroscopy of vibrational overtone bands: A proposal}
\author{Kevin K. Lehmann}
\email{lehmann@virginia.edu}
\affiliation{
 Departments of Chemistry and Physics, University of Virginia, Charlottesville VA, 22904-4319 }
\date{\today}
\begin{abstract}
This paper presents an analysis of near-resonant, ro-vibrational two-photon spectroscopy and the use of cavity ring-down spectroscopy for its detection.
Expressions are derived for the photon absorption rate of a three-level system, correct to all orders and the simpler expressions that result
from various approximations.   The analysis includes the angular momentum projection degeneracies and linear or circular polarization of the exciting field.  Expressions
are derived for the rate of two-photon power loss for light inside a resonant cavity.   Explicit calculations are made for excitation of the $\nu_3$ 
mode of $^{12}\textrm{C}^{16}\textrm{O}_2$ for which the two-photon excitation spectrum is dominated by a single $v_3 = 0 \rightarrow 2, Q(16)$ line at $\tilde{\nu} = 2335.826$\,cm$^{-1}$.
This transition 
has an intermediate $v_3 = 0 \rightarrow 1,P(16)$ one-photon transition that is off resonance by  0.093 cm$^{-1}$ (2.8 GHz).
At 1\,torr total pressure, the Q(16) two-photon transition has a calculated cross-section of $2.99 \cdot 10^{-38}$\,cm$^4$s per CO$_2$ molecule 
in the $J = 16$ state or $2.24 \cdot 10^{-39}$\,cm$^4$s per CO$_2$ molecule at 300\,K is calculated.  Analysis of the sensitivity limits for 2-photon cavity
ring-down spectroscopy predicts a theoretical detection limit of 32\,ppq ($10^{-15}$) Hz$^{-1/2}$ for $^{12}\textrm{C}^{16}\textrm{O}_2$, higher sensitivity
than has been realized using one-photon absorption.   The analysis predicts that most polyatomic molecules will have sparse, Doppler-Free two-photon absorption
spectra, which will dramatically increase the selectivity of trace gas detection of samples with multiple components with overlapping absorption bands.
This is demonstrated by the predicted mid-IR two-photon absorption spectrum of butadiene using theoretical spectroscopic constants.
\end{abstract}
\maketitle

Trace gas detection has been revolutionized by the use of low-loss optical cavity enhanced spectroscopic methods.   Starting with the introduction of the Cavity Ring-down technique in 1988,\cite{okeefe88} many forms of cavity enhanced spectroscopies (CES) have been developed and used for a wide range of applications.\cite{CRDS-book09, CES-book14}   Most of these applications have involved detection of gases, though applications for liquids~\cite{Xu02,Qu13} and surfaces~\cite{Pipino97,Powell09, Sangwan16} also are common.   Essentially all prior work has involved one-photon spectroscopy, in either the linear or saturated absorption regime.   One of the most impressive achievements has been development of instruments to measure $^{14}$C$^{16}$O$_2$ in pure CO$_2$ samples,~\cite{Galli16} with a sensitivity of 5 ppq (mole fraction in parts per $10^{15}$) after 2 hours of integration, well below the $\sim 1$\, ppt fractional abundance of $^{14}$C for atmospheric CO$_2$.  Several scientific instrument companies specialize in CES based trace gas analyzers.  CES results in an effective absorption path-length that is $(1-R_{\rm M})^{-1}$ (where $R_{\rm M}$ is the mirror reflectivity) times larger than the physical length of the cell.  In the near-IR through most of the visible spectral range, mirrors with path-length enhancements of $\sim10^5$ are commercially available. 

In trace detection, both sensitivity and selectivity are key.  The use of CES allows for remarkable sensitivity, but in one-photon absorption, the selectivity is limited by the high density of weak transitions of most polyatomic molecules that leads to spectral overlap.  One can minimize that overlap by working at pressures where the transitions are Doppler-broadened, but that compromises sensitivity.  Also, given the high enhancement of the power inside the sample cavity compared to the output power, at low sample pressure one can have optical saturation of the detection transition(s) at cavity output powers that give signals on the detectors only modestly higher than detector noise.~\cite{Bucher00}  This is particularly problematic when doing sensing with Mid-IR radiation, where the molecular transitions are strong but the detectors have limited sensitivity.   The rate of absorption, in the case of saturation, is limited by spectral hole burning.  Optical saturation has an advantage in that the saturating part of the cavity loss does not give an exponential decay, which is exploited in the Saturated-Absorption Cavity Ring-Down Spectroscopy (SCARS) technique~\cite{Giusfredi10} to distinguish it from the nonsaturated contributions of cavity loss, though the two loss contributions remain highly correlated.~\cite{Lehmann14}
	
In this paper, I discuss  a novel approach to CES that utilizes near-resonant, degenerate ro-vibrational two-photon absorption (TPA) by the gas contained in a low loss optical cavity.   Degenerate TPA from counter-propagating optical fields is intrinsically Doppler-Free,~\cite{Vasilenko70, Cagnac73, Biraben74} therefore, regardless of thermal velocity, all molecules in the correct lower state will absorb in a transition with frequency width limited by the homogeneous width of the transition.~\cite{Vasilenko70}   The unsaturated, peak absorption strength of TPA will saturate as a function of pressure once the homogeneous width due to pressure broadening exceeds the transit time broadening, which is typically only $\sim$100 kHz for practical optical cavities.     See Biraben\cite{Biraben19} for a recent review of the early work in Doppler-Free two-photon spectroscopy.

\section{Three level optical Bloch Equation, with equal relaxation rates }

Traditionally, the theory of two-photon absorption has been treated using second order, time dependent perturbation theory.~\cite{Goppert-Meyer31, Mahr75, McClain77}
In the electric dipole approximation, the rate of two-photon excitation from state 1 to 3 by absorption of two-photons of angular frequency $ \omega = (E_3 - E_1 ) / 2\hbar $ is proportional to the square of an amplitude, $S^{(2)}_{13}$, which can be written in terms
of a sum over virtual intermediate states, 2,~\cite{McClain77} 
\begin{equation}
S^{(2)}_{13} = \sum_2 \frac{  \mu_{12} \mu_{23}    }{ E_2 - E_1 - \hbar \omega  + i \hbar \gamma_{12}}  \label{eq:S2_13}
\end{equation}
where $\mu_{12}$ and $\mu_{23}$ are the transition electric dipole moments between the three states projected on the polarization direction of the driving electric field 
and $\gamma_{12}$ is the dephasing rate for the $1 \leftrightarrow 2$ transition.  It is evident that
there is a resonance enhancement when a state $2$ has allowed transitions to both states 1 and 3 and has energy nearly half way between these states such that $E_2 - E_1 - \hbar \omega \approx 0$.
This work considers cases where one particular such state 2 dominates the sum for $S^{(2)}_{13}$.  It is possible for such a
system to be driven sufficiently hard that the perturbation treatment is no longer accurate.   One can derive steady state solutions
to the density matrix for such a driven three level system and these allows calculation of the photon absorption rate for arbitrary excitation conditions.
 
\begin{figure}[htbp]
\begin{center}
\includegraphics{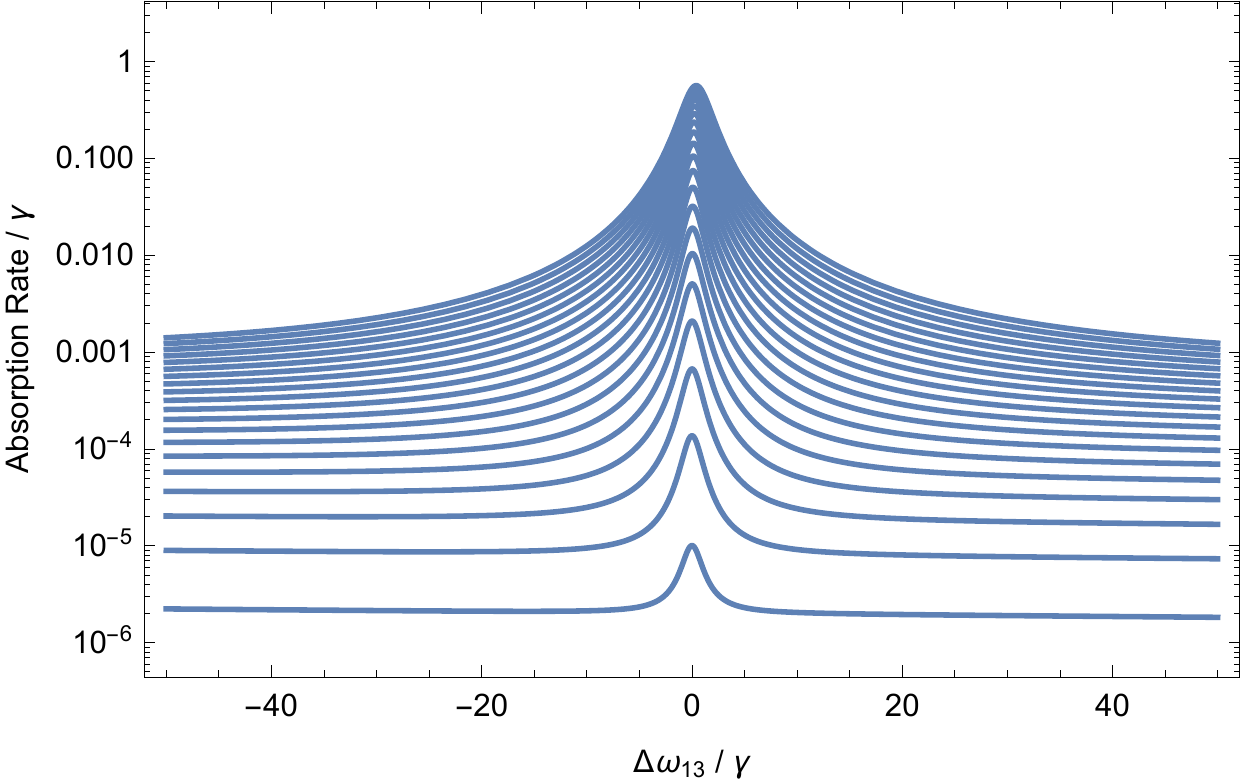}  
\caption{Steady-State photon absorption rate as a function of detuning, $\Delta \omega_{12}$, from the two-photon resonance angular frequency, both normalized by $\gamma$, the relaxation rate.
Calculations assumed that the detuning of the transition to the intermediate state (2) equals $1000  \gamma$, that $\Omega_{23} = \sqrt{2} \Omega_{12}$  and that $\Omega_{12}= \gamma ,2\gamma, \dots 20\gamma$. }
\label{fig:RateVsDw13}
\end{center}
\end{figure}

 Consider a three level system, with states labeled $1,2, 3$ with optical transitions between states $1 \leftrightarrow 2$ and $2 \leftrightarrow 3$ with definitions
 $\Omega_{12} = \mu_{12} {\cal E} / 2 \hbar$, $\Omega_{23} = \mu_{23} {\cal E} / 2 \hbar$ where $\cal E$ is the optical electric field amplitude at the position of the molecule,
  $\hbar \Delta \omega_{12} = E_2 - E_1 - \hbar \omega$ and $2 \hbar \Delta \omega_{13}  = E_3 - E_1 - 2 \hbar \omega$, where $E_i$ is the energy of state $i$.  
  Assume that at equilibrium only state $1$ is populated and all  population and dephasing relaxation rates are equal to $\gamma$.  
 This is typically a good approximation in ro-vibrational spectroscopy as relaxation is often dominated by inelastic collisions.
 Under these assumptions, along with the rotating wave approximation, the time evolution of the components of the dressed state density matrix is given by:~\cite{Oreg84}
\begin{eqnarray}
- \dot \rho_{11} &=& \gamma (\rho_{11} - 1) +i \Omega_{12}^* \rho_{12} - i \Omega_{12} \rho_{21} \nonumber \\
-\dot \rho_{22} &=& \gamma \rho_{22} -i \Omega_{12}^* \rho_{12} + i \Omega_{12} \rho_{21}  +i\Omega_{23}^*\rho_{23}-i \Omega_{23} \rho_{32} \nonumber \\
-\dot \rho_{33} &=& \gamma \rho_{33} -i \Omega_{23}^* \rho_{23} + i \Omega_{23} \rho_{32} \label{eq:drho} \\
-\dot \rho_{12} = -\dot \rho_{21}^* &=& i \Omega_{12} (\rho_{11} - \rho_{22} ) +  (\gamma - i \Delta \omega_{12} ) \rho_{12}   + i \Omega_{23}^* \rho_{13} \nonumber  \\
-\dot \rho_{13}  = -\dot \rho_{31}^* &=& i \Omega_{23} \rho_{12}  + ( \gamma - i \Delta \omega_{13} ) \rho_{13} - i \Omega_{12} \rho_{23}  \nonumber \\
-\dot \rho_{23} = - \dot \rho_{32}^* &=&  -i \Omega_{12}^* \rho_{13} + i \Omega_{23} (\rho_{22} - \rho_{33} ) + (\gamma + i (\Delta \omega_{12} - \Delta \omega_{13}))\rho_{23} \nonumber
\end{eqnarray}
The steady state rate of photon absorption per molecule is given by $ R_{\rm ss}  = \gamma ( \rho_{22} + 2 \rho_{33} )$. 
The full steady state solution for $R_{\rm ss}$ (derived using Mathematica) is
\begin{eqnarray}
&R_{\rm ss} &= 2 \gamma  \left| \Omega _{12} \right|^2 \left( \left| \Omega_{12} \right|^4 + 7 \left| \Omega_{23} \right|^4 + 2 \left| \Omega_{23} \right|^2 \left( \gamma^2 + 4 \left| \Omega_{23} \right|^2 + \left( \Delta \omega_{12} - \Delta \omega_{13} \right) \Delta \omega_{13} \right) \right. \nonumber \\
&+& \left.  \left( \gamma^2 + \left( \Delta \omega_{12} - \Delta \omega_{13} \right)^2 \right) \left(  \gamma^2 + \Delta \omega_{13}^2 \right)  
+ \left[ \Omega_{23} \right|^2 \left( 8 \gamma^2  + 2 \Delta \omega_{12}^2 - 3 \Delta \omega_{12} \Delta \omega_{13} + 6 \Delta \omega_{13}^2 \right) \right)  /  \nonumber \\
&& \left[   4 \left| \Omega_{12} \right|^6 + \left| \Omega_{12} \right|^4 \left( 9 \gamma^2 + 12 \left| \Omega_{23} \right|^2 + \Delta \omega_{12}^2 + 8 \Delta \omega_{12} \Delta \omega_{13} - 8 \Delta \omega_{13}^2 \right) \right.  \nonumber \\
&+& \left( \gamma^2  + 4 \left| \Omega_{23} \right|^2 + \left( \Delta \omega_{12} - \Delta \omega_{13} \right)^2 \right) \times \nonumber \\
&& \left(\left| \Omega_{23} \right|^4 + 2 \left| \Omega_{23} \right|^2 \left( \gamma^2 - \Delta \omega_{12} \Delta \omega_{13} \right) + \left(  \gamma^2 + \Delta \omega_{12}^2 \right)   \left(  \gamma^2 + \Delta \omega_{13}^2 \right)  \right) \nonumber \\
&+& 2 \Omega_{12}^2 \left( 6 \left| \Omega_{23} \right|^4 + 3 \gamma^2 \left( \gamma^2 + \Delta \omega_{12}^2 \right) + \Delta \omega_{12} \Delta \omega_{13} \left( \Delta \omega_{12}^2 - 3 \gamma^2 \right)
+ \left( 3 \gamma^2 + \Delta \omega_{12}^2 \right) \Delta \omega_{13}^2  \right. \nonumber \\
&-&\left.  \left.  4 \Delta \omega_{12} \Delta \omega_{13}^3 + 2 \Delta \omega_{13}^4 + \left| \Omega_{23} \right|^2 \left( 9 \gamma^2 + \Delta \omega_{12}^2 - \Delta \omega_{12} \Delta \omega_{13} + 10 \Delta \omega_{13}^2 \right) \right) \right]  \label{eq:full_rate}
\end{eqnarray}
which is rather formidable.   
Figure~\ref{fig:RateVsDw13}
displays plots of $R_{\rm ss}$ normalized by $\gamma$ as a function of $\Delta \omega_{13} / \gamma$ calculated with $\Delta \omega_{12} = 1000 \gamma,\, \Omega_{23} = \sqrt{2} \Omega_{12}$, and 
$\Omega_{12}  = \gamma, 2\gamma \dots 20 \gamma$.  The assumption $\Omega_{23} = \sqrt{2} \Omega_{12}$ is the double harmonic oscillator approximation (DHOA)\cite{Bernath_book}  in the case of two-photon excitation of a vibrational mode from the ground to the $v = 2$ state.  The DHOA
uses an expansion around equilibrium, up to quadratic for the potential and linear for the dipole moment.
Assuming $ |\Delta \omega_{12}| >> |\Omega_{12}|, |\Omega_{23}| >> |\Delta \omega_{13}|, \gamma$, and keeping the terms with the highest powers of $\Delta \omega_{12} $ in the numerator and denominator results in
\begin{equation}
R_{\rm ss} = \frac{ 4 \gamma |\Omega_{12} \Omega_{23} |^2 }{ \Delta \omega_{12}^2 \left( \gamma^2 + \Delta \omega_{13}^2 \right) + \left(  |\Omega_{12}|^2 + |\Omega_{23}|^2    \right)^2 } \label{eq-off-resonance}
\end{equation}
The rate is centered on $\Delta \omega_{13} = 0$, with power broadened half-width, half maximum equal to $\gamma \sqrt{1 + \left(  \left(  |\Omega_{12}|^2 + |\Omega_{23}|^2 \right) / \Delta \omega_{12} \gamma \right)^2 }$.  Neglecting the saturation correction, \textit{i.e.} assuming $ |\Omega_{12}|^2 + |\Omega_{23}|^2 <<  |\Delta \omega_{12}| \gamma $ , this is the same as was derived using perturbation theory by Vasilenko \textit{et al.}~\cite{Vasilenko70}   In this limit and on-resonance, $R_{\rm ss}(\Delta \omega_{13} = 0) \rightarrow 4 |\Omega_{12} \Omega_{23} |^2 / \gamma \Delta \omega_{12}^2$.  This is the excitation rate per molecule in the laser beam and the rate of loss of optical energy is proportional to $R_{\rm ss}$ times the number density of absorbers .  If the dephasing rate, $\gamma$ arises from collisions, as is typical in molecular ro-vibrational spectroscopy, then the rate of optical energy loss will be independent of sample density.    This pressure independence will break down when the pressure is sufficiently low that transit time broadening of the molecules exceeds $\gamma$, \textit{i.e.} that the mean free path of absorbers exceeds the size of the optical mode.

If we drive exactly on the two-photon resonance $\Delta \omega_{13} = 0$, we can write $R$
\begin{eqnarray}
R_{\rm SR} &=& 
  \frac{4 \gamma \left| \Omega _{23}\right|^2 \left| \Omega_{12}\right|^2 \left( \left| \Omega_{12} \right|^2+ \left| \Omega_{23}\ \right|^2+\gamma ^2 \right)}{  \left(\left| \Omega_{12} \right|^2+ \left| \Omega_{23} \right|^2 \right) \left( \left( \left| \Omega_{12} \right|^2+ \left| \Omega_{23}\right|^2+ \gamma^2\right)^2+ \gamma ^2 \Delta
   \omega_{12}^2 \right)} 
   \nonumber \\
     &&-  \frac{2 \gamma  \left( \left| \Omega_{23} \right|^2 -\left| \Omega_{12} \right|^2\right) \left| \Omega_{12} \right|^2}{ \left(  \left|
   \Omega_{12} \right|^2+ \left| \Omega_{23} \right|^2 \right) \left( 4 \left( \left| \Omega_{12} \right|^2+ \left| \Omega_{23} \right|^2  \right) +\gamma^2 + \Delta \omega _{12}^2 \right)  } \nonumber \\
  &\stackrel{\Omega_{23} = \sqrt{2} \Omega_{12}}{\longrightarrow}&
  \frac{2}{3} \gamma |\Omega_{12}|^2  \left( \frac{ 4 \left( \gamma^2 + 3 |\Omega_{12}|^2     \right)  }{  \left(  \gamma^2 + 3 |\Omega_{12}|^2    \right)^2   + \gamma^2 \Delta \omega_{12}^2 }
- \frac{1}{   \gamma^2  + 12 |\Omega_{12} |^2 + \Delta \omega_{12}^2}      \right)
    \label{eq:R_sr} 
\end{eqnarray}
which is the difference of two Lorentzians in $\Delta \omega_{12}$.
Saturation of the two-photon transition becomes important at an intensity when  $ |\Omega_{12}|^2 + |\Omega_{23}|^2 \sim  \left| \Delta  \omega_{12} \right| \gamma$.  When $\left| \Delta  \omega_{12} \right|^2 >>  |\Omega_{12}|^2 , |\Omega_{23}|^2  >> \left| \Delta  \omega_{12} \right| \gamma$,  the photon absorption rate saturates at $R_{\rm SR} \rightarrow 4 \gamma \left(   \frac{  |\Omega_{12} \Omega_{23} | }{   |\Omega_{12}|^2 + |\Omega_{23}|^2 }  \right)^2$ which in the DHOA $\rightarrow (8/9) \gamma$.  This can be compared to a steady state absorption rate of $\gamma$ assuming hard saturation with $\rho_{11} = \rho_{33} = 1/2$, expected when the TPA rate is low compared to $\Delta \omega_{12}$, or with $\rho_{11} = \rho_{22} = \rho_{33} = 1/3$,  expected when the TPA rate is large compared to $\Delta \omega_{12}$.

$R_{\rm SR}$ can be expanded in powers of $\Omega_{12}$ and $\Omega_{23}$ to obtain the absorption rate in powers of intensity or photon number density.   Keeping the terms up to fourth power in the $\Omega_{ij}$'s, \textit{i.e.} quadratic in light intensity, the low power limit ia
\begin{equation}
R_{\rm SR}  = \frac{2 \gamma  \Omega_{12}^2}{\gamma^2+\Delta \omega_{12}^2}
 -\frac{8 \gamma  \Omega_{12}^4}{\left(\gamma^2+\Delta \omega_{12}^2\right)^2}+\frac{4 \Omega_{23}^2 \Omega_{12}^2}{\gamma  \left(\gamma^2+
 \Delta \omega_{12}^2\right)}+ \ldots \label{eq:R_srExp}
 \end{equation}
The first term on the right is the linear absorption from one-photon $1 \rightarrow 2$ absorption, the second the leading saturation term of that transition, and the last the TPA $1  \rightarrow 3$.  This last term will dominate if $\left| \Omega_{23} \right| >> \gamma$.

\begin{figure}[htbp]
\begin{center}
\includegraphics[width=15cm]{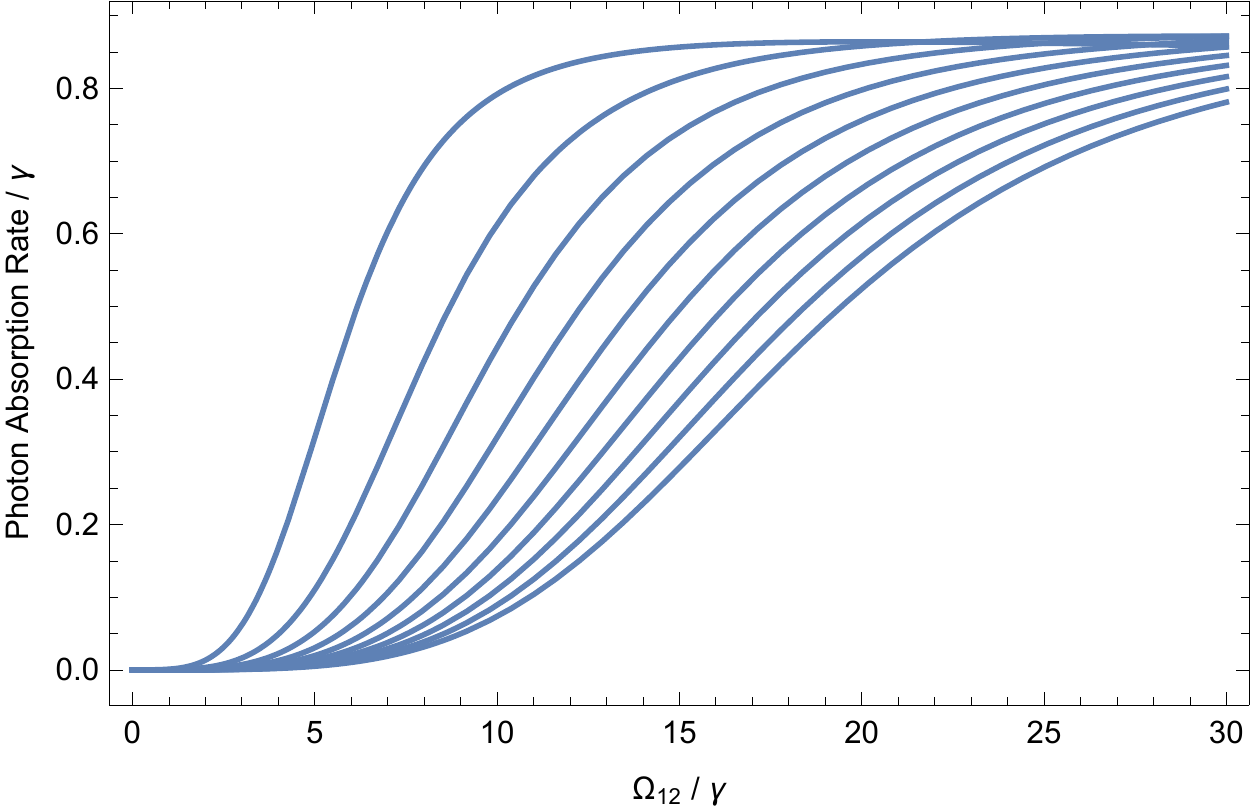}  
\caption{Steady-State photon absorption rate at the two-photon resonance as a function of $\Omega_{12}$, both normalized by $\gamma$.
The rates for values of $\Delta \omega_{12} = 100 \gamma , 200\gamma , \dots 1000\gamma $ are given by the curves from left to right in the figure.  
 $\Omega_{23} = \sqrt{2} \Omega_{12}$ (DHOA) and that 
 the Doppler broadening of the one-photon absorption is sufficiently smaller than the detuning
that it can be neglected have been assumed. }
\label{fig:RateVsDw12}
\end{center}
\end{figure}

\begin{figure}[htbp]
\begin{subfigure}[b]{ \textwidth}
\centering
\caption{Total Photon Absorption Rate}
\includegraphics[width = 0.5 \textwidth]{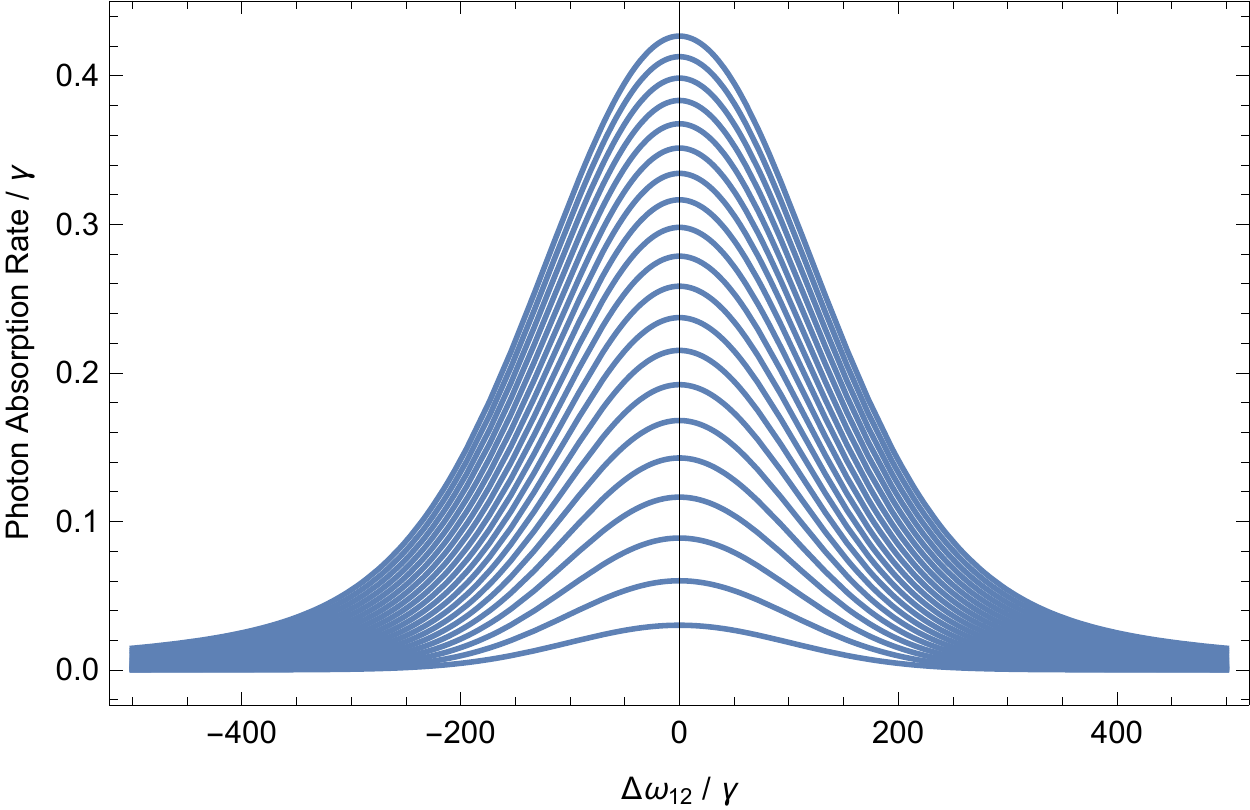}  
\end{subfigure}
\begin{subfigure}[b]{\textwidth}
\caption{One-Photon Absorption Rate}
\includegraphics[width = 0.5 \textwidth]{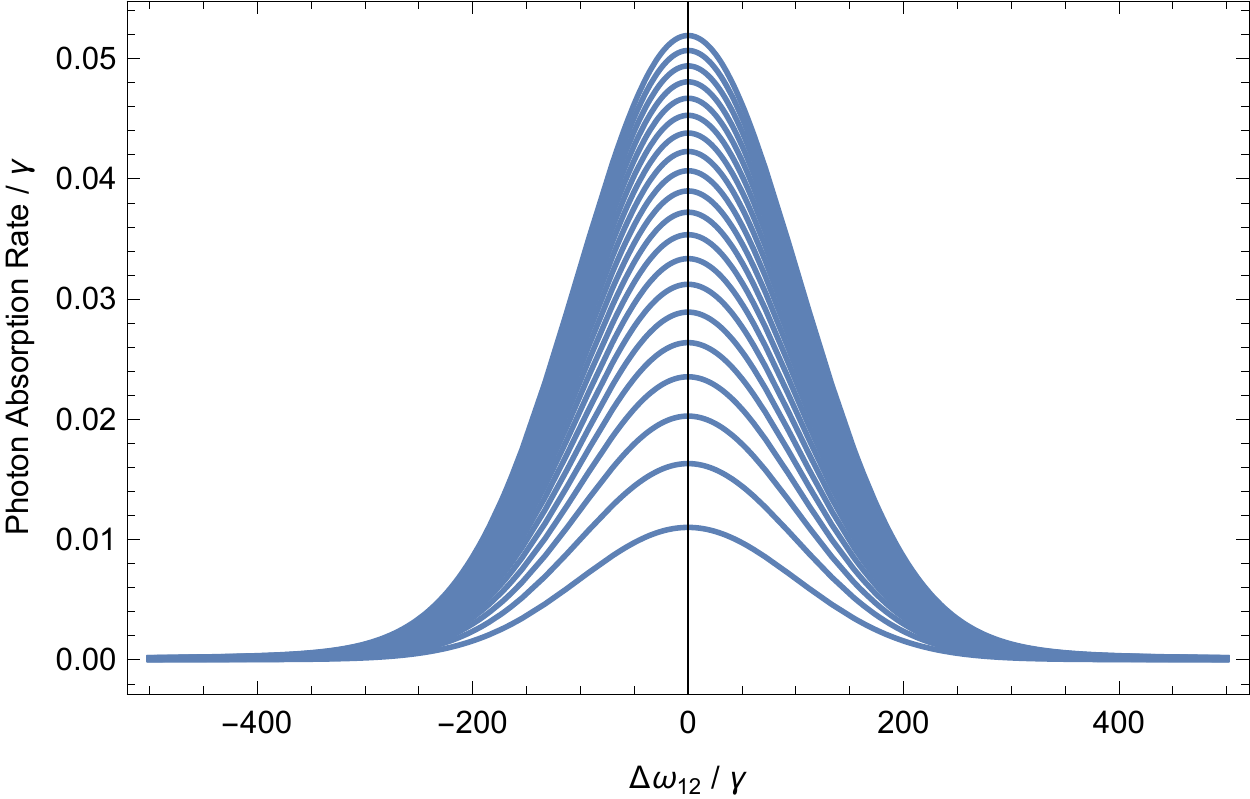}  
\end{subfigure}
\caption{a) Steady-State photon absorption rate at the two-photon resonance as a function of the one-photon detuning, $\Delta \omega_{12}$, normalized by $\gamma$.  A Doppler broadening standard deviation of $100 \gamma$ and $\Omega_{23} = \sqrt{2} \Omega_{12}$ are assumed.   Curves are plotted as a function of $\Omega_{12}^2 = \gamma^2, 2\gamma^2, \ldots, 20\gamma^2$, going from bottom to top.  b) The Steady-State photon absorption rate with the same parameters except $\Omega_{23} = 0$, \textit{i.e.} without the second transition.} 
\label{fig:RateVsDw12D}
\end{figure}

The above expressions should be applicable in the limit that $|\Delta \omega_{12}|$ is much larger than the Doppler width for that one photon transition.
For $|\Delta \omega_{12}|$ comparable or less than the Doppler width, the TPA rate given by Eq.~\ref{eq:R_sr}  convoluted with a normalized Gaussian Doppler distribution function for
$\Delta \omega_{12}$  gives a photon absorption rate per
molecule in terms of the normalized Voigt lineshape function~\cite{vandehulst47}  $g_{\rm V}( \Delta \omega_{12}, \sigma_{\rm D}, \gamma )$ where $\sigma_{\rm D}^2 = k_B T_g  \omega_{12}^2  / M c^2$
with $T_g$ the translational temperature of the absorbers, and $M$  the molecular mass of the analyte.  This gives a photon absorption rate:
\begin{eqnarray}
R_{\rm SR}& =& \frac{2 \pi \left| \Omega_{12} \right|^2}{  |\Omega_{12}|^2 + |\Omega_{23}|^2 } \left( \left| \Omega_{23} \right|^2 g_{\rm V} \left( \Delta \omega_{12}, \sigma_{\rm D},
\frac{\gamma^2 + \left| \Omega_{12} \right|^2 + \left| \Omega_{23} \right|^2   }{ \gamma   } \right) \right. \nonumber \\
  && -  \left. \frac{ \gamma \left( |\Omega_{23}|^2 - |\Omega_{12}|^2 \right)}{ \sqrt{\gamma^2 + 4 |\Omega_{12}|^2 + 4 |\Omega_{23}|^2 }  } 
g_{\rm V} \left( \Delta \omega_{12}, \sigma_{\rm D},   \sqrt{\gamma^2 + 4 |\Omega_{12}|^2 + 4 |\Omega_{23}|^2 }   \right) \right) \nonumber  \\
R_{\rm SR} &\underset{|\Omega_{23}|^2 = 2  |\Omega_{12}|^2}{\rightarrow}&   
 \frac{ 8 \pi   \left| \Omega_{12} \right|^2}{3} g_{\rm V} \left(\Delta \omega_{12} ,  \sigma_{\rm D} ,  \frac{ \gamma^2 + 3 \left| \Omega_{12} \right|^2  }{  \gamma }          \right)  \nonumber \\
 && - \frac{ 2 \pi \gamma   \left| \Omega_{12} \right|^2 }{  3 \sqrt{ \gamma^2 + 12 \left| \Omega_{12} \right|^2  } }  g_{\rm V} \left(\Delta \omega_{12} ,  \sigma_{\rm D} , \sqrt{ \gamma^2 + 12 \left|  \Omega_{12} \right|^2   } \right)
\end{eqnarray}
Note here, the second term is negative, canceling part of the contribution of the first term.
Figure~\ref{fig:RateVsDw12} displays the two-photon resonance photon absorption rate as a function of $\Omega_{12}$
for a range of values for the one-photon detuning, assuming the DHOA.
Figure~\ref{fig:RateVsDw12D} plots the photon excitation rate per molecule as a function of $\Delta \omega_{12}$, both normalized  by $\gamma$ and assuming a Doppler broadening
parameter $\sigma_{\rm D} = 100 \gamma$.   Curves are plotted (from bottom to top) for $\Omega_{12}^2  =  \gamma^2, 2 \gamma^2, \ldots 20 \gamma^2$.
The upper panel is the rate with both one and TPA ($\Omega_{23} = \sqrt{2} \Omega_{12}$), the lower where there is no TPA ( $\Omega_{23} = 0$ ).  Under the assumed parameters, even at exact simultaneous one and two-photon resonance, opening up the two-photon resonance increases the rate of photon absorption by nearly an order of magnitude. This is due to the fact that the one photon resonance is Doppler-broadened and the one photon excitation burns a hole in the ground state velocity distribution, while the entire Doppler profile can absorb two-photons.

Convoluting the normalized Doppler line shape with the power series expansion for $R_{\rm SR}$, Eq.~\ref{eq:R_srExp}, gives
\begin{eqnarray}
R_{\rm SR} &=& 2 \pi |\Omega_{12}|^2 \left(  1 + 2 \frac{ |\Omega_{23}|^2 - |\Omega_{12}|^2 }{\gamma^2}     \right) g_{\rm V} \left( \Delta \omega_{12}, \sigma_{\rm D}, \gamma \right)  \nonumber \\
&& + 
\frac{4 \pi |\Omega_{12}|^4}{\gamma} \left(  \frac{ \partial g_{\rm V} \left( \Delta \omega_{12}, \sigma_{\rm D}, \gamma \right) }{ \partial \gamma  }  \right) + \ldots
\end{eqnarray}

If $|\Delta \omega_{12}|$ is large compared to $\sigma_{\rm D}$, the sum of the two $\Omega_{12}^4$ terms is smaller than the $\Omega_{12}^2 \Omega_{23}^2$ term by a factor of $2 ( |\Omega_{12}| \gamma / |\Omega_{23}| \Delta \omega_{12})^2 = (\gamma/  \Delta \omega_{12})^2$ in the harmonic approximation,
and thus will be negligible at low pressures.   The small value of the $\Omega_{12}^4$ contribution is due to near equality of $\gamma \partial g_{\rm V}/\partial \gamma$ and $g_{\rm V}$ in this limit.

\section{Influence of Spatial Degeneracy}

Up to this point, I have only considered three level systems without the spatial angular momentum projection degeneracy associated with levels with nonzero total angular momentum $J$, \textit{i.e.} the $2J+1$ fold $M$ degeneracy.
For linear optical polarization, it is convenient to take the axis of angular momentum projection quantization parallel with the optical electric field, which results in optical absorption transitions having a $\Delta M = 0$ selection rule.  For circular polarization,
taking the axis of quantization along the direction of optical propagation, results in a $\Delta M = \pm 1$ selection rule for absorption from left or right handed polarized light respectively.   Due to the change in handedness of the light upon normal reflection, the forward and backward propagating waves pump the same sign of $\Delta M$. The signs are flipped for stimulated emission.   Linear absorption
rates are independent of optical polarization state for an isotropic sample, which has equal initial population in each $M$ state for any fixed set of values for the other quantum numbers.  Although TPA 
rates are different for linear and circular polarization,~\cite{Bonin84}  they are the  same for right and left polarized radiation, so we will only consider polarization states ($p$) as linear ($L$) and circular ($C$).   Changes in TPA with polarization can be used to assign symmetries of the states involved.~\cite{McClain71, McClain77}

Each of the $\Omega_{i,j}$ will have a dependence on the $M$ quantum numbers that can be represented as $\Omega_{i,j} (M_i, M_j ) = \Omega_{i,j} \, \phi_p (J_i,M_i, J_j, M_j)$ where $\phi_p$ are
the direction cosine matrix elements for polarization state $p$.  The nonzero values are given in Table~\ref{direction_cosine}.~\cite{Townes55}  The results presented above continue to hold if we replace $\Omega_{12}$
by $\Omega_{12} \, \phi_L (J_1, M, J_2, M)$ and $\Omega_{23}$ by $\Omega_{23} \phi_L ( J_2, M, J_3, M )$ for linear polarization and by $\Omega_{12} \phi_C (J_1, M, J_2, M+1)$ and $\Omega_{23}$ by $\Omega_{23} \phi_C ( J_2, M+1, J_3, M+2 )$ for the case of circular polarization and then average the rate over the $2J_1 + 1$ initial values of $M$.  Because the $\phi$ terms are different for the
fundamental and overtone transitions, the DHOA simplification can only be used if the H{\"o}nl-London factors\cite{Bernath_book} for the two transitions are nearly equal. 

For the photon absorption rate, neglecting saturation, we can replace the factor $\Omega_{12}^2 \, \Omega_{23}^2$ by
\begin{equation}
\Omega_{12}^2 \, \Omega_{23}^2 \rightarrow  \frac{  \Omega_{12}^2 \, \Omega_{23}^2}{2 J_1 + 1} \sum_{M_1} \phi_p (J_1, M_1, J_2, M_2)^2 \phi_p (J_2, M_2, J_3, M_3)^2 
\end{equation}
Using the line strength factor, $S$, Einstein spontaneous emission rate, $A_{i \rightarrow j}$, optical intensity, $I$, optical electric field amplitude, $\cal E$, transition frequency, $\nu$,, 
\begin{eqnarray}
S_{ij} &=& 3 \left< v_i, J_i || \mu || v_j, J_j \right>^2 \sum_{M_1, M_2} \phi_p (J_1, M_1, J_2, M_2)^2 = \frac{3 \pi \epsilon_0 \hbar c^3 (2 J_j +1)}
{\omega_{ij}^3 }  A_{i \leftarrow j } \\
&\Omega_{ij}^2&(J_1, M_1, J_2, M_2) = \frac{ {\cal E}^2 \left< v_i, J_i || \mu || v_j, J_j \right>^2}{4 \hbar^2}  \phi_p (J_1, M_1, J_2, M_2)^2 
 \end{eqnarray}
Using $I = \epsilon_0  c \, |{\cal E}|^2 /2 = h \nu N_p$, the effective value for $\Omega_{12}^2 \Omega_{23}^2$ can be written as
\begin{eqnarray}
\Omega_{12}^2 \Omega_{23}^2 &=& \frac{\pi^2 c^4 I^2}{4 \hbar^2 \omega^6} A_{1,2} A_{2,3} \,  a_p(J_1, J_2, J_3) \label{eq:EffOmega4}\\
a_{\rm L} (J_1, J_2, J_3) &=&  \frac{ (2 J_2 +1)(2 J_3 +1)  }{ 2 J_1 +1   } \frac{ \sum_{M} \phi_L(J_1,M,J_2, M)^2 \cdot \phi_L(J_2, M, J_3, M)^2}{  \sum_{M} \phi_L(J_1,M,J_2, M)^2 \cdot \sum_{M} \phi_L(J_2,M,J_3, M)^2  }  \\
a_{\rm C} (J_1, J_2, J_3) &=&  \frac{ (2 J_2 +1)(2 J_3 +1)  }{ 2 J_1 +1   } \frac{ \sum_{M} \phi_C(J_1,M,J_2, M+1)^2 \cdot \phi_0(J_2, M+1, J_3, M+2)^2}{  \sum_{M} \phi_C(J_1,M,J_2, M+1)^2 \cdot \sum_{M} \phi_X(J_2,M,J_3, M+1)^2     } \nonumber 
\end{eqnarray}
The dimensionless factors $a_p(J_1, J_2, J_3)$ account to the spatial degeneracy.  Their nonzero values are given in Table~\ref{aValuesTable}.  All the $a_p$ factors are on the order of unity, but some two-photon transitions are favored by linear polarization (RR, RP, PR, PP, QQ) and some by circular polarization (QR, QP, RQ, PQ).  For the $P(J) - R(J-1)$ sequence, linear polarization produces four times the excitation rate as circular polarization.

\begin{table}[htp]
\caption{Direction Cosine Matrix Element Factors~\cite{Townes55} }
\begin{center}
\begin{tabular}{|c|c|c|c|c|c|c|c|}
\hline
 &  J' = J-1  &  J' = J   & J' = J+1\\
\hline
$\phi_L(J,M,J', M)$ &     $ \frac{ \sqrt{J^2 - M^2}}{ (J^2  ( 4 J^2 - 1))^{1/4}}$        & $\frac{M}{\sqrt{J(J+1) }}$                                            &  $\frac{ \sqrt{(J+1)^2 - M^2}}{ ((J+1)^2  (2 J+1) (2 J+3))^{1/4}}$  \\ \hline
$\phi_C(J,M, J',M+1$) & $\frac{\sqrt{(J-M) (J-M-1)}}{ ( 4 J^2  ( 4 J^2 - 1))^{1/4}}$         &  $ \frac{ \sqrt{(J-M)(J+M+1) }   }{ \sqrt{2J(J+1) }} $    & $ -\frac{ \sqrt{(J+M+1) J+M+2)   }}{ (4(J+1)^2  (2 J+1) (2 J+3))^{1/4} }$ \\ \hline
\end{tabular}
\end{center}
\label{direction_cosine}
\end{table}

\begin{table}
\caption{ $a_p$ factors to correct for $M$ dependence in off resonance two-photon excitation rate    }
\begin{center}
\begin{tabular}{|c|c|c|c|c|c|}
\hline
$J_2$     & $J_3$  &  $a_L(J_1, J_2, J_3)$ & $a_C(J_1, J_2, J_3)$ & $a_L$ high J  & $a_C$ high J \\ \hline \hline
$J_1+1$ & $J_1+2$   & $\frac{6 (2 J_1 +5)   }{ 5 ( 2 J_1 + 1)   }$ & $\frac{ 9 (2 J_1 +5)   }{ 5 (2 J_1 +1)   } $                                                         & 6/5 & 9/5  \\ \hline
$J_1+1$ & $J_1+1$   & $\frac{3 J_1 (2 J_1 +3)   }{ 5 (J_1+1)( 2 J_1 + 1)   }$ & $\frac{ 9 J_1  (2 J_1 +3)   }{ 10 (J_1+1) (2 J_1 +1)   } $           &3/5  &9/10    \\ \hline
$J_1> 0$ & $J_1+1$   & $\frac{3 (J_1+2)(2J_1+3)   }{ 5 (J_1+1)( 2 J_1 + 1)   }$ & $\frac{ 9 (J_1+2)  (2 J_1 +3)   }{ 10 (J_1+1) (2 J_1 +1)   } $ &3/5 & 9/10   \\ \hline
$J_1 + 1$ & $J_1$   & $\frac{3( 4 J_1^2 + 8J_1 +5)   }{ 5 (J_1+1)( 2J_1 + 1)   }$ & $\frac{ 3 J_1  (2 J_1 -1)    }{ 10 (J_1 +1)( 2J_1 +1)   } $       &6/5 & 3/10    \\ \hline
$J_1 $& $J_1$           & $\frac{3( 3 J_1^2 + 3J_1 -1)   }{ 5 J_1( J_1 + 1)   }$ & $\frac{ 3 (2 J_1 +3)  (2 J_1 -1)    }{ 10 J_1 ( J_1 +1)   } $             &9/5 & 6/5    \\ \hline
$J_1 -1 $ & $J_1 > 0$   & $\frac{3 (4 J_1^2 + 1)   }{ 5 J_1 (  2J_1 + 1)   }$ & $\frac{ 3 (J_1+1)  (2 J_1 +3)    }{ 10 J_1 ( 2J_1 +1)   } $                &6/5  & 3/10  \\ \hline
$J_1 >0  $ & $J_1 -1> 0$   & $\frac{3 (J_1-1)(2J_1-1)   }{ 5 J_1 (  2J_1 + 1)   }$ & $\frac{ 9 (J_1-1)  (2 J_1 -1)    }{ 10 J_1 ( 2J_1 +1)   } $         & 3/5  & 9/10\\ \hline
$J_1 -1 >0$ & $J_1 -1$   & $\frac{3 (J_1+1)(2J_1-1)   }{ 5 J_1 (  2J_1 + 1)   }$ & $\frac{ 9 (J_1+1)  (2 J_1 -1)    }{ 10 J_1 ( 2J_1 +1)   } $         &  3/5 &  9/10 \\ \hline
$J_1-1 >0$ & $J_1-2$   & $\frac{6 (2 J_1 -3)   }{ 5 ( 2 J_1 + 1)   }$ & $\frac{ 9 (2 J_1 -3)    }{ 5 (2 J_1 +1)   } $                                                      &   6/5& 9/5\\ \hline
\end{tabular}
\end{center}
\label{aValuesTable}
\end{table}

\section{Influence of standing wave and Doppler Broadening of the transitions of the intermediate state}

The expression given in Eq~\ref{eq:EffOmega4} is applicable for a single, traveling wave.   To realize Doppler-free TPA, a pair of counter propagating waves is necessary.   If both waves have intensity $I$ the TPA amplitude will be double because of the equal contributions of two paths, differing in the order of absorption from the two waves.   For both of these amplitudes, the $\exp( i k (r(0) + v t))$ and $exp(- i k (r(0) + v t))$ factors (with $\pm k$ the wavevector for the two traveling waves making up the standing wave) will cancel, leading to a Doppler-free absorption since the amplitude is independent of velocity.   Another way to arrive at the factor of two is to use a Van Vleck transformation to eliminate the intermediate state 2.   In that case, also assuming that $|\Delta \omega_{12}| >> |\Delta \omega_{13}|$, the effective off-diagonal matrix element coupling states 1 and 3 is 
\begin{eqnarray}
\Omega_{13}^{\rm eff} &=& \frac{\Omega_{12} ( \exp( i k r(t)) + \exp( - i k r(t))) \Omega_{23} ( \exp( i k r(t)) + \exp( - i k r(t))) }{\Delta \omega_{12} +i \gamma }  \nonumber \\
&=& \frac{\Omega_{12} \Omega_{23}  \left( 2 + \exp( 2i k r(t)) + \exp( -2 i k r(t))) \right)}{\Delta \omega_{12} +i \gamma } \label{eq:Omega13_eff}
\end{eqnarray}
 The term independent of $r(t)$ gives the Doppler free TPA and the two terms dependent on $r(t)$ give TPA Doppler shifted by $\pm k v$, and thus Doppler-broadened transitions.   Since the steady state TPA rate is proportional to the square of $\Omega_{13}^{\rm eff}$, the steady-state Doppler free TPA is four times larger than given in Eq~\ref{eq:R_srExp}.  The Doppler-broadened terms will give steady state rates twice Eq~\ref{eq:R_srExp} times the normalized Doppler distribution function,
 $g_{\rm D}(\omega)$. On resonance, the Doppler Free TPA will dominate over the Doppler-broadened TPA by a factor of $2 g_{\rm H} / g_{\rm D}$, which will be very large when the homogeneous width is much less than the Doppler width.  For the Doppler free TPA from a standing wave with intensity $I$ in each direction, Eq~\ref{eq:EffOmega4} needs to be replaced by
 \begin{equation}
\Omega_{12}^2 \Omega_{23}^2 = \frac{\pi^2 c^4 I^2}{\hbar^2 \omega^6} A_{1,2} A_{2,3} \,  a_p(J_1, J_2, J_3) \label{eq:EffOmega42}
\end{equation}

Keeping the $ \Omega_{12}^2 \Omega_{23}^2 $ term in Eq~\ref{eq:R_srExp}, and using Eq.~\ref{eq:EffOmega42} for the effective value, the two-photon resonance steady state absorption rate is
\begin{equation}
R_{\rm SR} = \frac{ 4  \Omega_{12}^2 \Omega_{23}^2 }{\gamma \left( \gamma^2 + \Delta \omega_{12}^2  \right)} =   \frac{4 \pi^2 c^4 }{ \hbar^2 \omega^6 \gamma \left( \gamma^2 + \Delta \omega_{12}^2  \right)    }  I^2 A_{1,2} A_{2,3} \,  a_p(J_1, J_2, J_3)
\end{equation}
It has been assumed that $\omega_{12}$ and $\omega_{23}$ can be replaced by the two-photon angular frequency, $\omega = (\omega_{12} + \omega_{23})/2$.

Averaging over the Doppler Broadening of $\Delta \omega_{12}$ gives
\begin{eqnarray}
R_{\rm SR} &=&  \frac{4\pi^3 c^4 }{ \hbar^2 \omega^6 \gamma^2    } g_{\rm V}\left(\Delta \omega_{12}, \sigma_{\rm D}, \gamma    \right) A_{1,2} A_{2,3} \,  a_p(J_1, J_2, J_3)  I^2 \\
& = & \frac{4\pi^3 c^4 }{ \omega^4 \gamma^2    } g_{\rm V}\left(\Delta \omega_{12}, \sigma_{\rm D}, \gamma    \right) A_{1,2} A_{2,3} \,  a_p(J_1, J_2, J_3)  N_{\rm p}^2
\end{eqnarray}
where $N_{\rm p}$ is the photon flux rate.   Converting molecular parameters, except the $A_{i \rightarrow j}$ values,  to dimensions of cm$^{-1}$, as that is how spectroscopic 
parameters are usually tabulated, produces
\begin{eqnarray}
R_{\rm SR} &=&  \frac{1 }{ 16 \pi^4 h^2 c^5 \tilde{\nu}^6 \Delta \tilde{\nu}_{\rm H}^2    } g_{\rm V}\left(\Delta \tilde{\nu}_{12}, \tilde{\sigma}_{\rm D}, \Delta \tilde{\nu}_{\rm H}    \right) A_{1,2} A_{2,3} \,  a_p(J_1, J_2, J_3)  I^2  \\
& = & \sigma^{(2)}_{13} N_{\rm p}^2 =  \frac{1 }{ 16 \pi^4 c^3 \tilde{\nu}^4 \Delta \tilde{\nu}_{\rm H}^2    } g_{\rm V}\left(\Delta \tilde{\nu}_{12}, \tilde{\sigma}_{\rm D}, \Delta \tilde{\nu}_{\rm H}    \right) A_{1,2} A_{2,3} \,  a_p(J_1, J_2, J_3) N_{\rm p}^2 \label{eq:sigma2} \\
\sigma^{(2)}_{13}  &=&2.38 \cdot 10^{-35} {\rm cm}^4 {\rm s} \, \frac{  g_{\rm V}\left(\Delta \tilde{\nu}_{12}, \tilde{\sigma}_{\rm D}, \Delta \tilde{\nu}_{\rm H}    \right) A_{1,2} A_{2,3} \,  a_p(J_1, J_2, J_3)  }
{\tilde{\nu}^4 \Delta \tilde{\nu}_{\rm H}^2  } \nonumber \\
 & \underset{|\Delta \tilde{\nu}_{12}| >> \tilde{\sigma}_{\rm D}, \Delta \tilde{\nu}_{\rm H} }{\rightarrow}&
 7.58 \cdot 10^{-36} {\rm cm}^4 {\rm s} \, \frac{   A_{1,2} A_{2,3} \,  a_p(J_1, J_2, J_3)  }{\tilde{\nu}^4 \Delta \tilde{\nu}_{12}^2  \Delta \tilde{\nu}_{\rm H}   }
   \nonumber
\end{eqnarray}
where for the last two expressions, $A_{i,j}$ in s$^{-1}$, $g_{\rm V}$ in cm, and all wavenumber quantities in cm$^{-1}$ is assumed.

The homogeneous broadening wavenumber can be calculated as $\Delta \tilde{\nu}_{\rm H} = \gamma/ 2\pi c = b_p P_{\rm g}$ where $b_p$ is the pressure broadening coefficient (half width at half maximum, HWHM) and $P_{\rm g}$ the
pressure of the gas sample.
The above results are rates and cross sections per molecules in state $1$.  For reference to the molecules of a specific compound in the gas, multiply by the fraction of those
molecules in state $1$, $f_1 = g_1 \exp \left ( - h c T_1 / k_{\rm B} T_g \right) / Q(T_g)$, where $g_1$, and $T_1$, are the degeneracy and term value (energy converted to cm$^{-1}$) of state 1
respectively and $Q(T_g)$ is the partition function for that molecule.   If $|\Delta \tilde{\nu}_{12}| >> \sigma_D$ and $\Delta \tilde{\nu}_{\rm H}$, then $g_{\rm V} \rightarrow \Delta \tilde{\nu}_{\rm H} / (\pi \Delta \tilde{\nu}_{12}^2)$.  If $ \tilde{\sigma}_{\rm D} >> |\Delta \tilde{\nu}_{12}|$ and $\Delta \tilde{\nu}_{\rm H}$, \textit{i.e.} there is exact double resonance for a frequency near the center of the Doppler lineshape, then $g_{\rm V} \rightarrow 1/\sqrt{2 \pi} \sigma_D$.

Based upon Eqs.~\ref{eq-off-resonance} and \ref{eq:EffOmega4}, it can be shown that, in the off-resonance case ( $\Delta \tilde{\nu}_{12} >> \tilde{\sigma}_{\rm D},
|\Omega_{12}|, |\Omega_{23}|$ and $\gamma$), optical saturation will reduce the on-resonance TPA by a factor of
$1 + (I/I_{\rm{sat}})^2$,  where
\begin{equation}
I_{\rm sat} = \frac{8 \pi^3 h c^3  \Delta \tilde{\nu}_{12} \, \Delta \tilde{\nu}_{\rm H}  \,  \tilde{\nu}^3   }{  (A_{12} + A_{23}) \sqrt{a_{\rm p}}  } \label{eq:Isat}
\end{equation}
Note that the saturation power scales linearly with $\Delta \tilde{\nu}_{\rm H} $ and thus pressure, unlike for a Doppler-broadened line where the saturation power scales as the square of $\Delta \tilde{\nu}_{\rm H}$ and thus pressure squared.   Also, in the high power limit, $I >> I_{\rm sat}$, the TPA rate approaches a constant while
for one-photon absorption the excitation rate is proportional to $\sqrt{I}$ in high intensity limit, as long as $|\Omega_{12}|$ remains small compared to the Doppler
width of the line.

\section{Cavity Ring-Down detection of two-photon absorption}

To relate the steady state TPA rate to the rate of decay of intra-cavity intensity, 
let the number density of molecules in state $1$ at equilibrium be $N_1$ and assume the molecules are excited by the TEM$_{00}$ mode of a optical cavity with one way intra-cavity
power $P_{\rm ic}$.  If the cavity intensity decay rate and the transit rate of molecules across the cavity mode are slow compared to $\gamma$, the molecular absorption remains in steady state with the instantaneous intensity at the position of each molecule.  
The rate at which optical energy is absorbed from the intra-cavity field will equal the integral of $h c \tilde{\nu}  R_{\rm ss} N_1$ over the cavity. 
Assuming the cavity is nearly confocal with length $L$, the integral of $I^2$ over the cavity volume gives $(\pi/2) \tilde{\nu} P_{\rm ic}^2$.~\cite{Yariv75}
Neglecting saturation, the power absorbed by the molecules inside the cavity is given by:
\begin{equation}
P_{\rm abs} = \frac{g_{\rm V}\left(\Delta \tilde{\nu}_{12}, \tilde{\sigma}_{\rm D}, \Delta \tilde{\nu}_{\rm H} \right) A_{12} A_{23} \, a_p(J_1,J_2, J_3) N_1 f_1} {128 \pi^3 h c^4 \tilde{\nu}^4 \left(\Delta \tilde{\nu}_{\rm H}^2 +   \Delta \tilde{\nu}_{13}^2     \right) } P_{\rm ic}^2
\end{equation}
Note that this is independent of the cavity length because a shorter cavity has a more tightly focused mode. The optical energy stored in the cavity equals $(2L/c) P_{\rm ic}$, so $dP_{\rm ic}/dt = -(c/2L) P_{\rm abs}$.  

Consider on resonance excitation on the TPA ($ \Delta \tilde{\nu}_{13} = 0$), and an ideal gas at pressure and temperature, $P_{\rm g}, \, T_{\rm g}$, for which $ \Delta \tilde{\nu}_{\rm H} = b_{\rm p} P_g $ and 
$N_1 = (P_g / k_{\rm B}\,T_{\rm g})  x_{\rm a}$
where $x_{\rm a}$ is the mole fraction of the analyte in the sample gas.  These assumptions lead to a rate of loss of intra-cavity power
\begin{eqnarray}
\frac{d P_{\rm ic}}{dt} &=& -\gamma_1 P_{\rm ic} -\gamma_2 P_{\rm ic}^2  \label{eq:diff} \\
\gamma_1 &=&  c \left(\alpha + \frac{1-R_{\rm m}}{L} \right)  \\
\gamma_2 &=&   \frac{ g_{\rm V}\left(\Delta \tilde{\nu}_{12}, \tilde{\sigma}_{\rm D}, b_p P \right) A_{12} A_{23} \, a_p(J_1,J_2, J_3)  x_a f_1} {256 \pi^3 h c^3 L k_{\rm B}  T_g \tilde{\nu}^4 b_p^2 P_g }    \label{eq:gamma2} \\
\gamma_2 & & \underset{|\Delta \tilde{\nu}_{12}| >> \tilde{\sigma}_{\rm D}, b_p P}{\rightarrow}  \frac{ A_{12} A_{23} \, a_p(J_1,J_2, J_3)  x_a f_1} {256 \pi^4 h c^3 L k_{\rm B}  T_g  \Delta \tilde{\nu}_{12}^2 \tilde{\nu}^4 b_p}  \\
&=& 1.64 \cdot 10^{19} \,({\rm W\,s})^{-1}  \times \frac{ A_{12} A_{23} \, a_p(J_1,J_2, J_3)  x_a f_1} {L  T_g  \Delta \tilde{\nu}_{12}^2 \tilde{\nu}^4 b_p} \nonumber
\end{eqnarray}
where the last expression assumes $A_{ij}$ in s$^{-1}$, $L$ in cm, T$_{\rm g}$ in K, $b_p$ in cm$^{-1}$ atm$^{-1}$, and wavenumber quantities in cm$^{-1}$.
Eq.~\ref{eq:diff} can be integrated to give the Power vs. time
\begin{equation}
P_{\rm ic}(t) = \frac{\gamma_1 P_{\rm ic}(0) \exp(-\gamma_1 t) }{ \gamma_1 + \gamma_2 P_{\rm ic}(0) \left(  1 -   \exp(-\gamma_1 t)  \right)    }  \label{eq:2PhotonTransient}
\end{equation}
Above,  $\alpha$ denotes the linear extinction coefficient of the intracavity sample, including the one photon $1 \rightarrow 2$ transition, and $R_{\rm m}$ is the geometric mean of the power reflectivity of the two cavity mirrors.
Note that for a detuning that is large compared to the Doppler width, the unsaturated TPA rate is independent of pressure while when the intermediate state detuning is well within the Doppler broadening, the TPA rate is inversely proportional to gas pressure.  This will break down if the pressure is so low that either saturation or transit time broadening of the TPA cannot be ignored.  

Previously, I published expressions to allow the simultaneous determination of $\gamma_1$ and $\gamma_2$ by a least squares fit of an observed cavity
decay transient to Eq.~\ref{eq:2PhotonTransient}.~\cite{Lehmann14}, both in the detector and shot-noise limits.  This work also provided predicted standard 
errors for the fitting parameters.   For the case of shot-noise-limited detection, the standard error for $\gamma_2$ from a fit to a single decay can be written as:
\begin{equation}
\sigma'(\gamma_2) = \frac{1}{P_{\rm ic}(0)} \sqrt{ \frac{6 h c \tilde{\nu} }{Q_{\rm det} P_{\rm det}(0) }}  \gamma_1^{3/2}  \label{eq:gamma2sigma}
\end{equation}   
where $Q_{\rm det}$ is the quantum efficiency of the detector, $P_{\rm ic}(0)$ the intracavity power at the start of the decay, and $P_{\rm det}(0)$ the optical power
on the detector at the start of the decay.   The single decay standard error in the determination of analyte mole fraction, $x_a$ can be calculated using Eqs~\ref{eq:gamma2} and \ref{eq:gamma2sigma} and will scale as $L^{-1/2}$.   The number of decays one can observe at optimal single to noise is proportional to $\gamma_1$ and thus $L^{-1}$, and so the predicted sensitivity limit for fixed detection bandwidth is predicted to be independent of $L$.

The lowest order transverse mode of a stable optical cavity (TEM$_{00}$) will have a Gaussian shape normal to the optical axis, with a beam radius $w$ (distance off-axis where the magnitude of the electric field falls to $1/e$ of the on-axis field ). The on-axis intensity is given by $2 P_{\rm ic}/ (\pi w^2)$ and we can define a saturation power $P_{\rm sat} = (\pi w^2 / 2) I_{\rm sat}$.  Integration of $I^2$ perpendicular to the axis with and without saturation demonstrates that saturation will reduce the TPA of a perpendicular splice by a factor of $ (P_{\rm sat}/P_{\rm ic})^2 \ln \left( 1 + (P_{\rm ic}/P_{\rm sat})^2   \right)$.

\section{Application to Carbon Dioxide and other molecules}

Consider a parallel two-photon transition of a linear molecule with a $\Sigma$ ground state  $v_s = 0 \rightarrow 1 \rightarrow 2$.  There will be $S(J) (\Delta J = 2)$, $Q(J) (\Delta J = 0)$ and $O(J) (\Delta J = -2)$ transitions, with the largest resonance enhancement for one of the $Q$ branch transitions.  As the $v = 1 \rightarrow 2$ ``hot band'' is typically red shifted from the $v = 0 \rightarrow 1$ fundamental, the resonance will be largest for the $J$ values near where the $P(J)$ branch of the fundamental crosses the $R(J-1)$ branch of the hot band.   Neglecting centrifugal distortion and higher order vibration-rotation interaction constants, the nearest resonant two-photon transitions will be $Q(J)$ transitions with detunings given by:
\begin{equation}
\Delta \nu_{12} (v_s = 0, J \rightarrow v_s = 1, J-1 \rightarrow v_s = 2, J) = -X_{ss}  + 2(B_0 - \alpha_s ) J
\end{equation}
$X_{ss}$ is the diagonal anharmonicity constant of mode $s$, $B_0$ the ground vibrational state rotational constant, and $\alpha_s$ the vibration-rotation interaction constant for mode $s$.   
Typical values for the $X_{ss}$ constants are $-5$ to $-50$\,cm$^{-1}$. The transition with the smallest $|\Delta \nu_{12}|$ occurs for the integer $J$ value closest to $-X_{ss}/2(B_0 - \alpha_s )$ and this $J$ value has a detuning less than $B_0 - \alpha_s$, except in the cases of $J$ states with zero nuclear spin weight.  If the ground state is not $\Sigma$ symmetry, we also have additional near resonant transitions $J \rightarrow J-1 \rightarrow J-1$ and $J \rightarrow J \rightarrow J+1$ with detuning of $X_{ss}+B_0 J$ and $X_{ss} + (B_0 - 2 \alpha_s) (J+1)$ respectively.  While we can expect, on average, the closest of these to be even closer to two-photon resonance, the $Q$ branch transitions of parallel bands are weak, so these are unlikely to be the strongest TPA line unless they are particularly near resonant.  

Perpendicular modes of such a linear molecule, $t$, have possible transitions from the ground state to the $2 \nu_t, \Sigma (l = 0)$ and $\Delta (l = \pm 2)$ vibrational states.  For the $\Sigma$ state, there will be near resonances again for the $J \rightarrow J-1 \rightarrow J$ transition, but with detuning given by $ X_{tt} - g_{tt}/2 + 2(B_0 -\alpha_t - q_t)J$.  For the transitions to the $\Delta$ state, the $J \rightarrow J-1 \rightarrow J$ excitation path has detuning $ X_{tt} + g_{tt}/2 + 2(B_0 -\alpha_t - q_t/2)J$.  In addition, there are near resonant transitions with excitation paths $J \rightarrow J-1 \rightarrow J-1$ and $J \rightarrow J \rightarrow J+1$.  These have detuning of $X_{tt} + g_{tt}/2 + (B_0 - q_t/2)  J$ and $X_{tt} + (B_0 - 2\alpha_t + q_t)/2(J+1)$.  Unlike the case of the parallel transitions, the Q branch transitions are strong, and the closest resonance of these will have a detuning less than $\sim B_0 / 2$.

As a quantitative example, consider TPA by $^{12}$C$^{16}$O$_2$ gas in the region of the $\nu_3$ fundamental.  Transitions for both the fundamental and hot band in this mode are listed in the HITRAN database,~\cite{Gordon17} which gives the parameters needed to calculate the two-photon cross section.  $\nu_3$ is the antisymmetric stretching mode, well known to be extremely intense with $A_{01} =215.4$/s, $A_{12} = 401$/s (a ratio quite close to the DHOA prediction), while   $X_{33} = -12.63$\,cm$^{-1}$.  The  two-photon transition with the nearest resonant intermediate state is $Q(16)$ at $\tilde{\nu} = 2335.826$\,cm$^{-1}$, which is the mean of the transition wavenumber for the $P(16)$ of the fundamental band and $R(15)$ of the fundamental to overtone band.   The $P(16)$ transition of the fundamental band is off-resonance by only 0.093 cm$^{-1}$ (2.8 GHz).  The other excitation path that contributes to the $Q(16)$ two-photon amplitude is the R(16) of the fundamental and P(17) of the fundamental to overtone band, which  is off-resonance by $\sim 66 B \sim 25.7$\,cm$^{-1}$ with nearly the same dipole matrix elements.  Thus, it makes a negligible contribution to the TPA. The air broadening coefficients of each transition is listed in HITRAN as $b_{\rm p} = 0.076$\,cm$^{-1}$/atm ( 3 MHz/torr).   
At a gas temperature $T_g = 300$\,K, J=16 is near to the Boltzmann maximum with fractional population $f_1 = 0.075$ and the Doppler width parameter $\Delta \tilde{\sigma}_{\rm D} = 0.00186$ cm$^{-1}$ (56 MHz).  For this transition, $a_{\rm L}(J_0,J_1,J_2 )=1.165$.    At 1\,torr total pressure and $T = 300$],K , a two-photon cross section of $2.99 \cdot 10^{-38}$\,cm$^4$s per CO$_2$ molecule 
in the $J = 16$ state or $2.24 \cdot 10^{-39}$\,cm$^4$s per CO$_2$ molecule is calculated from these parameters using Eq.~\ref{eq:sigma2}.  This can be compared to a typical
two-photon cross section for electronic transitions, which are on the order of $10^{-50}$\,cm$^4$s, known as one G{\"o}ppert-Mayer (GM) unit.\cite{McClain77}  Even Squaraine Fluorophores, developed
to be ``Ultra-Bright" for two-photon microscopy~\cite{Podgorski12} have peak two-photon excitation cross sections on the order of $10^{4}$ GM, \textit{i.e.} $10^{-46}$\,cm$^4$s, 7 orders of magnitude smaller than
for the resonantly enhanced transition in CO$_2$.  Using Eq.~\ref{eq:Isat}, a two-photon saturation power is calculated as 4.93 kW/cm$^2$ times the gas pressure in torr, which can be compared to a saturation power of 1.3 W/cm$^2$ times the square of the pressure in torr for the one-photon P(16) line of CO$_2$.

For the cavity ring-down signal estimate, we will take the parameters for the cavity used by Galli \textit{et al.}~\cite{Galli16} in their $^{14}$CO$_2$ SCARS measurements:   $L = 1$\,m, $T_{\rm M} = 87$\, ppm, $1-R_{\rm M} = 190$\,ppm, which imply $\gamma_1 = 5.7 \cdot 10^{4}$/s, neglecting any linear loss from the sample.  At the CO$_2$ TPA wavenumber, the beam radius at the center of the cavity is $w_0 = 825\,\mu$m. 
For mode-matched, monochromatic excitation of such a cavity, the predicted peak transmission is $(T_{\rm M}/(1-R_{\rm M}))^2 = 0.21$.  Assuming we use a laser with 100\,mW power to excite the cavity, the predicted output power is $P_0 = 21$\,mW; an initial intracavity power $P_{\rm ic}(0) = 241$\,W each way, for which the peak intracavity intensity is 22.5\,kW/cm$^2$.  For these parameters, $\gamma_2 = 2.13
 \cdot 10^{8} $x(CO$_2$)/(W\,s) is calculated.
Taking a detector quantum efficiency  $Q_{\rm det} =0.63$ and initial power on the detector at the start of a decay as $P_{\rm det}(0) = 10$\,mW, \ and $\gamma_1 = 5.69 \cdot 10^4$/s, Eq.~\ref{eq:gamma2sigma} gives $\sigma'(\gamma_2) = 3.75 \cdot 10^{-4}$/W\,s. 
Assuming cavity decays detected at a rate of 3\,kHz ($\sim \gamma_1 / 20$), a fit standard error for $\gamma_2$ of $6.85 \cdot 10^{-6}$ / W s $\sqrt{ \rm Hz}$ is calculated using Eq.~\ref{eq:gamma2sigma}.  
Comparing this to the calculated 
value of $\gamma_2$ as a function of x(CO$_2$) gives a predicted standard error in the determination of $x$(CO$_2$) = 32\,ppq ($10^{-15}$) Hz$^{-1/2}$.
This sensitivity value can be compared to the sensitivity limit of 5 ppq after 2 hr integration for $^{14}$C$\,^{16}$O$_2$ reported by Galli \textit{et al.}~\cite{Galli16} using the SCAR technique~\cite{Giusfredi10} and a cavity cell cooled to 170\,K and filled to 12\,mbar, 
which translates to a sensitivity of 425 ppq per $\sqrt{\rm Hz}$.  Their gas density corresponds to a 300\,K sample at 16 torr, for which the
two-photon saturated power is predicted as 79 kW/cm$^2$, 3.5 times the intracavity power.   With $P_{\rm ic} / P_{\rm sat} = 0.28$, a 3\% reduction of
the TPA due to optical saturation is calculated.

\begin{figure}[htbp]
\begin{center}
\includegraphics[width=17cm]{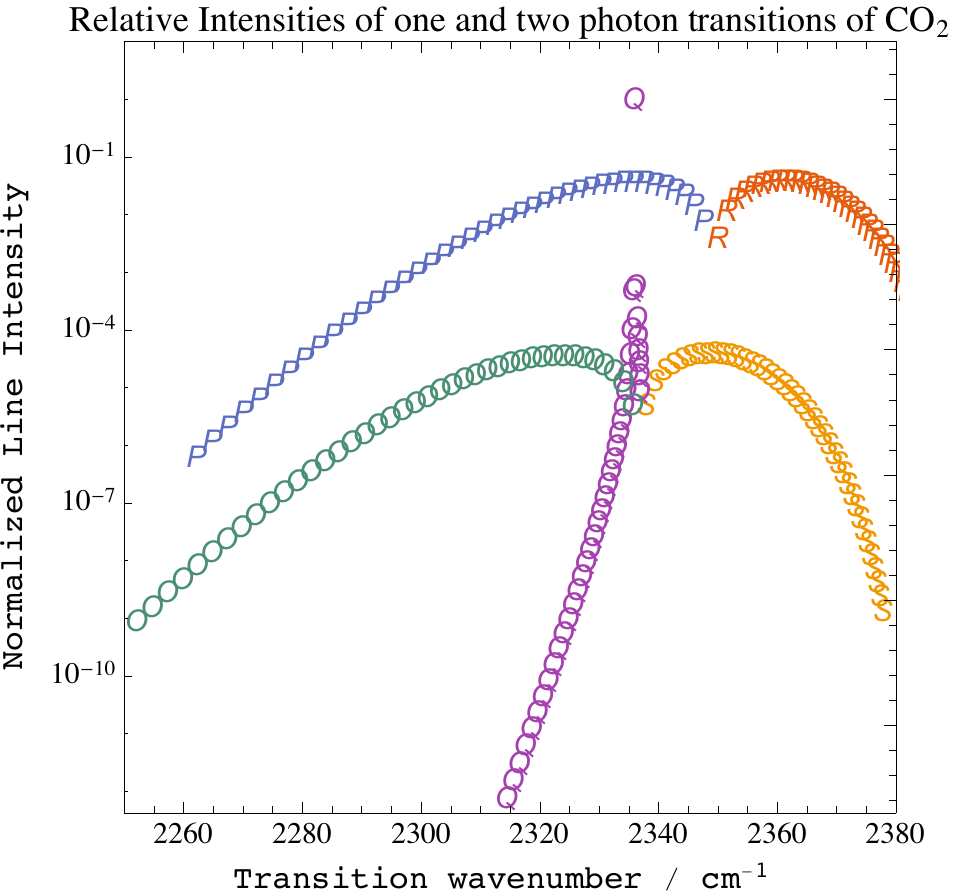}  
\caption{Comparison of relative intensities of one and two-photon transitions of $^{12}$C$\,^{16}$O$_2$.  Type of transition is indicated by letter with O, P, Q, R, S 
representing $\Delta J = -2, -1, 0, 1, 2$.   The one and two-photon bands are each normalized to have unity integrated intensity for the band }
\label{fig:CO2_one_vs_two}
\end{center}
\end{figure}

The selectivity of TPA is illustrated by looking at the relative strength of the different two-photon transitions in the $\nu_3$ fundamental region.  
Figure~\ref{fig:CO2_one_vs_two} displays a comparison on the calculated $^{12}$C$^{16}$O$_2$ $\nu_3 $ and $2\nu_3$ relative intensities for the different rotational transitions, normalized so that the sum of intensities for both one-photon (P and R branches), and the two-photon (O, Q, S branches) transitions equals unity.  Calculations used a 300\,K rotational temperature.
The $Q(16)$ transition is $\sim 250$ times stronger than the next most intense transitions ($ Q(14)$ and $Q(18)$ ) and about 1000 times stronger than the strongest lines in the O ($\Delta J = -2$) and S ($\Delta J = +2$) branches.  We can expect qualitatively similar behavior in the TPA spectrum of most molecules that could interfere in the detection of other analytes.  

The HITRAN online\cite{HITRAN_Online} database was searched for molecules for two-photon transitions where the fundamental and overtone level are listed, the later by either the corresponding hot band or by an overtone band from the ground state.   In the former, the $A$ values listed for both transitions that make up a resonant TPA triplet of states were used to calculate the TPA cross section of each transition.  In the latter case, the A value for the $v = 1 \rightarrow 2$ transition was estimated by using the DHOA.  The strongest transitions for each isotopic species are given in supplementary material to this paper.  

Now consider application of TPA to a relatively heavy asymmetric top molecule, which has much more complex level structure than CO$_2$.  For a molecule without symmetry, most ground vibrational state rotational levels will have 25 TPA transitions for each final vibrational level~\cite{McClain77}  ($O, P, Q, R, S$ branches for both $\Delta J$ and $\Delta K$ ),  even without considering transitions allowed by asymmetry mixing that gives intensity to $|\Delta K_{a,c}|  > 1$ transitions.   While sensitivity remains important, a more frequent limitation in the detection of these molecules, particularly using IR spectroscopy, is the difficulty in selectively detecting one species from another, particularly if they contain the same generic IR ``functional groups''.  

As another illustration of the sparseness of TPA spectra, I have calculated the TPA spectrum of trans-Butadiene (t-BD), point group $C_{2h}$, using anharmonic spectroscopic constants (kindly supplied to me by Dr. Manik Pradhan), using Gaussian-16, with the 6-311++g(3df,2p) basis set and the B3YLP functional.\cite{Maithani19}   Table~\ref{C4H6_constants_table} contains the calculated spectroscopic constants for the fundamentals and first overtone for each of the IR active normal modes of t-BD.  The predicted constants explicitly include near resonant Fermi and Darling Dennison resonances.  Normal modes numbered 10-13 are $A_u$ symmetry representations and give C-axis polarized IR fundamentals.   Modes numbered 17-24 are $B_u$ representations and have mixed A,B-polarized IR fundamentals.   Lacking the relative contributions of the A and B axis transition dipole matrix elements, they were assumed to be equal.   Calculations used the rigid rotor approximation, including all transitions with total angular momentum quantum number, $J$, up to 100.  At 300\,K, t-BD has a rotational partition function of 32\,037.   For fundamental bands, the strongest individual transitions have fractional intensities of (6.29, 4.83, 3.04, and 9.47) $\times 10^{-4}$ of the entire IR band for A, B, A/B hybrid, and C polarized bands, respectively, where the hybrid band is assumed to have equal A and B contributions.   One way to characterize an effective number of transitions is by the ratio $(\sum I_t)^2 / \sum I_t^2$ where $I_t$ is the absorption intensity of the $t$-th transition.   Neglecting the frequency factors in the intensities, the effective number of transitions is 3410, 7380, 9549, and 6420 for the A, B, A/B hybrid, and C polarized bands.  

The TPA spectrum of t-BD from the ground vibrational state to the two quanta level in each of the the IR allowed modes was calculated using Eq.~\ref{eq:S2_13} for the two-photon amplitude and expressing the intensity as proportional to the square magnitude of the transition amplitude times the lower state population.   The sum over intermediate states included all allowed  transitions in the fundamental level.   In order to approximately account for Doppler broadening of intermediate state resonances without calculating the Voigt function for each transition, the Doppler width was treated as homogeneous broadening, i.e. $\gamma = 2 \pi c \Delta \nu_D$ in Eq.~\ref{eq:S2_13}  where $\Delta \nu_D$ is the Doppler half width, half maximum.   300\,K thermal populations were assumed for the ground vibrational state and thermal population of the excited vibrational states was neglected.  Intensities of each two-photon transition were initially calculated using only the direction cosine matrix elements to allow comparison of different normal modes without the complication of their different IR intensity values, giving transition intensities in units of cm$^2$.  These are summarized in Table~\ref{C4H6_TPA_table}. The total intensity, summed over all TPA transitions is given in the table, and is found to vary widely with normal mode, spanning the range from 30 - 11,000\,cm$^2$.   The differences are due to the small differences in upper state rotational constants and the different values for the diagonal anharmonicity calculated from the $v=0 \rightarrow 1$ and $v = 1 \rightarrow 2$ vibrational intervals.   The effective number of transitions, as defined above for the fundamentals, varies between 2.74 and 67, demonstrating that the TPA spectra are dramatically sparser than then the one photon spectra.  

Another way to characterize how sparse the TPA spectra are is by the fraction of total intensity carried by the strongest transition in the band.  These vary from a low of 0.053 to a high of 0.59, i.e. 59\% of the total band intensity is contained in a single transition.   The rotational quantum numbers for the strongest transitions of each band are listed and are distinct in each case, reflecting the fact that the strongest transitions are extremely sensitive to the constants of the three vibrational states involved.   Also listed is the two-photon cross section for each of the strongest transitions (integrated over transition wavenumber) to give units of cm$^3$\,s.  
Given the high sensitivity to spectroscopic constants, the presently presented calculations are unlikely to be quantitatively accurate, but they should be considered representative of the expected size of these quantities, at laboratory temperatures, for a molecule of the size of t-BD.   

\begin{table}
\caption{Calculated Spectroscopic constants for trans-Butadiene vibrational states  \\ }
\begin{center}
\begin{small}
\begin{tabular}{|c|c|c|c|c|c|c|c|c|c|c|c|c|c|c|c|}
\hline
state&Sym&G(v)&Intensity&A&B&C \\
&&cm$^{-1}$&km/mol.&cm$^{-1}$&cm$^{-1}$&cm$^{-1}$ \\
\hline
G.S.&Ag&0&&1.3980866&0.1476468&0.1334412\\
\hline
v(10)&Au&1029.019	&22.44&1.394038&0.148405&0.133465\\
v(11)&Au& 927.810	&79.63&1.36601&0.147538&0.133479 \\
v(12)&Au& 530.252	&11.53&1.394707&0.147527&0.13344\\
v(13)&Au& 168.965	&0.60&1.338177&0.147815&0.133823 \\
v(17)&Bu&3080.483	&25.42&1.395197&0.147566&0.133357 \\
v(18)&Bu&2998.877	&3.87&1.39411&0.147563&0.133341 \\
v(19)&Bu&2977.928	&22.46&1.394491&0.147568&0.133342 \\
v(20)&Bu&1614.113	&26.66&1.395653&0.147366&0.133123 \\
v(21)&Bu&1388.271	&4.33&1.400366&0.147668&0.133366  \\
v(22)&Bu&1303.287	&0.68&1.403727&0.147775&0.133385 \\
v(23)&Bu& 995.753	&1.81&1.424255&0.146821&0.13336\\
v(24)&Bu& 303.799	&2.75&1.45607&0.147641&0.133369\\
\hline
&&G(v)&X(s,s) &A&B&C \\
&&cm$^{-1}$&cm$^{-1}$&cm$^{-1}$&cm$^{-1}$&cm$^{-1}$ \\
\hline
2v(10 )&Ag &2054.497&-3.541 &1.38999&0.149162&0.133491  \\
2v(11) & Ag &1857.59&1.97&1.333933&0.147428&0.133519 \\
2v(12 )&Ag&1058.985&-1.519&1.391328&0.147407&0.133439   \\
2v(13 )&Ag&336.738&-1.192 &1.278268&0.147982&0.134205  \\
2v(17)&Ag&6159.571&-1.395&1.392308&0.147483&0.133274   \\
2v(18 )&Ag&5999.975&2.221&1.390133&0.147479&0.133242    \\
2v(19 )&Ag &5995.06&39.204 &1.390896&0.147488&0.133244 \\
2v(20 )&Ag &3226.105&-2.121 &1.39322&0.147083&0.132806  \\
2v(21)&Ag&2773.766&-2.776&1.402646&0.147687&0.133292    \\
2v(22)&Ag &2596.266&-10.308&1.409368&0.147901&0.13333  \\
2v(23)&Ag&1992.229&0.723&1.450423&0.145993&0.13328  \\
2v(24)&Ag&609.207&1.609&1.514054&0.147635&0.133299 \\
\hline
\end{tabular}
\end{small}
\end{center}
\label{C4H6_constants_table}
\end{table}

\begin{table}
\caption{Calculated Two-Photon IR absorption of trans-Butadiene overtone of each IR allowed mode\\ }
\begin{center}
\begin{small}
\begin{tabular}{|c|c|c|c|c|c|c|c|c|c|c|c|c|c|c|c|}
\hline
mode&$\Delta \nu_{\rm D}$&total Int.&N eff&I max&$\nu$&$\int \sigma^2$& Transition & E'' \\
&MHz& cm$^2$ & & / I tot& cm$^{-1}$ &  cm$^3$ sec & & cm$^{-1}$ \\
\hline
10&18.9&146.8&10.49&0.1818&1019.9716&1.47819E-43&$26(4,22) \rightarrow 26(2,24)$&118.8634\\
11&17.1&205&2.79&0.5883&933.535&1.19819E-41&$35(3,32) \rightarrow 36(3,44)$&190.0674\\
12&9.8&372.9&2.75&0.4522&541.9089&3.09595E-42&$21(1,21) \rightarrow 23(3,21)$ &64.3248\\
13&3.1&1.10E04&8.864&0.1913&154.7277&1.59542E-41& $14(7,8) \rightarrow 14(5,10)$  &91.1398\\
17&56.6&52.96&14.91&0.1442&3074.0817&6.57796E-46&$18(3,16) \rightarrow 17(2,16)$	&59.4442\\
18&55.1&31.1&17.33&0.1331&2995.6409&9.19439E-48&$20(1,19) \rightarrow 19(0,19)$	&61.5281\\
19&54.7&30.26&36.18&0.0721&2986.1806&1.64691E-46&$26(13,14) \rightarrow 27(12,16)$	&311.2058\\
20&29.7&97.43&7.754&0.293&1604.6385&3.64311E-44&$35(3,33) \rightarrow 34(2,33)$	&188.7359\\
21&25.5&214.63&67.13&0.0529&1391.7118&6.7642E-46&$10(3,7) \rightarrow 10(3,7)$	        &17.0934\\
22&23.9&70.17&7.33&0.2747&1297.9526&3.73832E-47&$42(3,39) \rightarrow 42(3,39)$	&268.4731\\
23&18.3&117.62&8.72&0.2192&938.1758&1.29104E-45&$51(2,49) \rightarrow 49(3,47)$	&382.9722\\
24&5.6&97.19&4.87&0.4183&319.0226&3.53076E-43&$56(4,53) \rightarrow 58(3,55)$		&470.4185\\
\hline
\end{tabular}
\end{small}
\end{center}
\label{C4H6_TPA_table}
\end{table}

In many applications of Cavity ring-down and other forms of Cavity Enhanced Spectroscopy, the ultimate sensitivity limit is determined by drifts in the empty cavity loss.  By scanning the laser over a resonant absorption or by frequency jumping on and off resonance, cavity loss drift that changes slowly with wavelength relative to the absorption line-width are corrected for, but ``fringes'' that arise from interferences remain.  These can arise from scattering by any of the optics that couple back into the resonator.  The scattering of a single photon back into the cavity during the decay will just be detectable when the decay is shot-noise limited.~\cite{Huang11}  These interference fringes often have widths comparable with Doppler-broadened lines.  For the CO$_2$ fundamental, the Doppler FWHM is $\sim 130$\,MHz, similar to the FSR of an interference caused by a back reflection from an optic 1\,m from a cavity mirror (150\,MHz).  At total gas pressure on the order of 1\,torr, the TPA line-width will typically be on the order of a few MHz, which is far narrower than any fringe produced by feedback.  
Galli \textit{et al.}~\cite{Giusfredi10, Galli16} have demonstrated that the SCARS method (where the molecular absorption is saturated and thus changes during the cavity decay) allows a separation of the molecular loss from all the contributions to the empty cavity loss: only the latter is affected by the interference.  This is what allows them to improve their sensitivity limit by very long time integration.   This feature has also been theoretically demonstrated for TPA, with reduced correlation of the fitted linear and TPA loss rates compared to the saturable and non-saturable loss rates that are fitting parameters in the SCARS technique.~\cite{Lehmann14}  

To avoid strongly saturating the TPA the intracavity pressure should be selected such that the excitation rate per resonant molecule is less than the relaxation rate, which is proportional to the gas pressure.   For the CO$_2$ detection parameters assumed above, the two-photon excitation rate on-axis in the center of the cell will match the collisional dephasing rate for a gas pressure of 6.9 torr, where the pressure broadening of the transition will have a HWHM of 20 MHz.  Under the same conditions, the peak one way  power in the cell will be 362 times the saturation power of the fundamental P(18) line.

To efficiently enhance the intracavity power, the excitation laser should be locked to the cavity (or visa versa) with a relative frequency jitter of much less than the width of the cavity mode, the FWHM of equals $\gamma_1/2\pi = 9.5$ kHz for the assumed cavity length and loss.  Two locking schemes are considered.  Galli \textit{et al.}~\cite{Galli16}  used the Pound-Drever-Hall method~\cite{Drever83} to lock a Quantum Cascade laser to their cavity with a residual laser-cavity frequency jitter well below the cavity mode width, thus demonstrating that this is technically feasible even while observing cavity decay events which interrupt the stabilization.   They used an acquisition rate 2500 decays per second, which is close to the 3\,KHz assumed in the above calculation.   In the SCAR technique, the cavity loss must be numerically integrated to produce the predicted intensity decay curve, while TPA, without saturation,  produces an analytical intensity decay curve.   Thus the computational resources required to process a TPA decay is well below that for an equivalent SCARs decay transient. 
An alternative approach to efficient coupling is to use optical feedback from the cavity to spontaneously lock QCL laser to the cavity.~\cite{Dahmani87,Morville05,Kerstel06}  In this case, using a 3-mirror V-shaped cavity will eliminate the reflection from the input mirror; however, for identical mirror properties, doing so will reduce the intracavity power build up by a factor of four,~\cite{Morville05}
due to the loss of impedance matching.   This type optical locking will likely be more tolerant of vibrations and electronic noise than locking using electronic feedback and thus may be more promising for applications outside of a laboratory setting.

\section{Discussion}

The analysis presented in this paper has demonstrated that near-resonance, TPA of vibrational transitions, detected using a version of cavity ring-down spectroscopy, should be both highly selective and sensitive.  The selectivity comes about from its Doppler-Free resonances, sparse spectrum of transitions highly enhanced by  near resonances, and its unique temporal cavity decay transients which allows separation of the TPA loss from all linear or saturated loss contributions from the cavity itself or other components of the gas mixture.   Surprisingly, at least to the author, the calculated strength of the two-photon transitions are sufficient, with near resonant enhancement, such that the sensitivity (in terms of calculated lowest detection level) is not compromised.  High power, broadly tunable, narrow linewidth, c.w. IR lasers, such as Quantum Cascade Lasers, and low-loss mid-IR mirrors have become commercially available, both of which enhance TPA.  Crystalline Mirror Solutions has recently introduced mirrors with crystalline layers which have a measured loss $(1-R_{\rm M}) = 190$\,ppm and $T_{\rm M} = 144$\,ppm near $4.5\,\mu$m, which predicts a one-way power gain of 4000, compared to 2400 used in the above calculations.~\cite{Heckl_pc}.   For molecules with more complex
and dense level structure than CO$_2$, especially heavy asymmetric tops, a few two-photon transitions with detuning within the Doppler width can be expected, leading to further  increase in the two-photon cross section relative to state resolved one-photon absorption. 

Why has this not been done already?  The overwhelming number of previous studies of two-photon spectroscopy have examined electronic transitions of atoms and molecules.   This is perhaps not surprising in that fully allowed electronic transitions have transition dipole moments $\sim 10^2$ times larger than the corresponding vibrational transitions.  In addition, high power pulsed lasers have been
more readily available in the visible spectral region.   There was one attempt to observe two-photon excitation of NO$_2$ using CRDS, but this was unsuccessful.~\cite{Romanini99}  The excitation region scanned in this study has extremely weak NO$_2$ one-photon transitions, the one photon CRDS absorption sensitivity was well below current level, and the fitting did not address the expected changes in cavity decay temporal shape, so it is difficult to judge the significance of this failure.  

In the early days of mid-IR laser spectroscopy, the only readily available laser sources where fixed frequency atomic lasers (such as the IR He-Ne laser) and line
tunable molecular lasers such as the CO$_2$, NNO, and CO lasers.   These had limited tuning (limited by Doppler widths of the laser gain transitions) but had relatively high power.
Relevant to the present paper, a number of IR double resonance transitions were observed, even without cavity enhancement, in molecules such as CH$_3$F~\cite{Bischel75a,Bischel75b,Bischel76b,Prosnitz78}  and NH$_3$\cite{Bischel75a, Bischel76a, Bischel76c,Bokor79, Jacobs76}  by Stark shifting the two-photon transitions into resonance with the fixed frequency lines of a CO$_2$ laser.  

\section{Supplementary Material}

The supplementary material contain tables that list the strongest ro-vibrational two-photon transitions of molecules, calculated using data from the HITRANonline~\cite{HITRAN_Online} spectral database.   Two photon cross sections are calculated for a pressure of 1 torr, using the air-broadening pressure broadening coefficients from the database.   The cross sections are listed for molecules in the specific initial state, for the specific isotopologue  by scaling by the thermal fractional population of that state at the HITRAN reference temperature, 293\,K, and per molecule of the specific chemical species by scaling by the fractional population of the isotopologue as given by database.   Files in common separated value format.   Separate files are given for diatomics, linear molecules, asymmetric, and symmetric tops.   

\section{Acknowledgements}

The author wishes to thank Dr. Manik Pradhan of the S.N. Bose National Center for Basic Sciences for providing him with the results of his recent electronic structure calculations on trans-Butadiene, which allowed a realistic calculation of the two-photon vibrational spectrum for this molecule.  He also wishes to acknowledge helpful discussions and comments by Dr. Joseph T. Hodges, of N.I.S.T.    The University of Virginia supported this work.

\bibliography{Two-photon_Spectroscopy}

\end{document}